\documentclass[showpacs,showkeys,superscriptaddress]{revtex4-2}
\usepackage{amsmath}
\usepackage{amssymb}
\usepackage{graphicx}
\usepackage[hypertex]{hyperref}

\begin{document}
\title{Rapidly rotating Dirac stars
 }

\author{
Vladimir Dzhunushaliev
}
\email{v.dzhunushaliev@gmail.com}
\affiliation{
Department of Theoretical and Nuclear Physics,  Al-Farabi Kazakh National University, Almaty 050040, Kazakhstan
}
\affiliation{
Institute of Nuclear Physics, Almaty 050032, Kazakhstan
}
\affiliation{
Academician J.~Jeenbaev Institute of Physics of the NAS of the Kyrgyz Republic, 265 a, Chui Street, Bishkek 720071, Kyrgyzstan
}

\author{Vladimir Folomeev}
\email{vfolomeev@mail.ru}
\affiliation{
Institute of Nuclear Physics, Almaty 050032, Kazakhstan
}
\affiliation{
Academician J.~Jeenbaev Institute of Physics of the NAS of the Kyrgyz Republic, 265 a, Chui Street, Bishkek 720071, Kyrgyzstan
}

\author{Nassurlla Burtebayev}
\affiliation{
Institute of Nuclear Physics, Almaty 050032, Kazakhstan
}

\begin{abstract}
Within general relativity, we construct sequences of rapidly rotating Dirac stars consisting of a spinor fluid
described by an effective equation of state.
We find the physically relevant domain of stable configurations and calculate their principal
characteristics which are completely determined by the central density of the spinor fluid, the mass of the nonlinear spinor field,
and the velocity of rotation. It is demonstrated that for a certain choice of the spinor field mass, the main physical characteristics of the Dirac stars
are close to those that are typical of rotating neutron stars.
\end{abstract}

\maketitle

\section{Introduction}
Strongly gravitating configurations supported by various types of matter have been the object of
vigorous investigations in different formulations of the problem. The physical nature of such objects may vary considerably,
and is determined both by a type of matter supporting them and by a theory of gravity within which their modeling is performed.
 A typical object of investigations are neutron stars~-- compact configurations possessing characteristic masses of the order of 1-2 solar masses
 and radii 10-20 km~\cite{HPY}. In studying such objects,
the following aspects determining their physical characteristics are of great importance:
(i)~the choice of equation of state of matter creating such configurations;
(ii)~ the inclusion of rotation and the presence of strong magnetic fields; (iii)~a theory of gravity used in modeling the stars.
All these aspects are intensively studied in the literature by starting, first, with various physically reasonable assumptions about the nature of neutron stars and, second,
with matching theoretical models to the current observational data.

However, it is not impossible that other types of compact strongly gravitating objects do exist in nature. These can be, in particular, hypothetical boson stars
supported by scalar (spin-0) fields.
Their investigations are carried out in the same directions as those of neutron stars.
In doing so, depending on the specific parameters of the scalar field, the physical characteristics of such objects 
may vary over a wide range, from those that are typical of atoms to those which characterize galaxies~\cite{Schunck:2003kk,Liebling:2012fv}.

Apart from boson stars, a consideration of configurations consisting of fields with nonzero spins is also possible. Such hypothetical systems may be supported
by massive vector (spin-1) fields (Proca stars~\cite{Brito:2015pxa,Herdeiro:2017fhv,Sanchis-Gual:2019ljs,Dzhunushaliev:2021vwn}) and massless vector fields (Yang-Mills systems~\cite{Bartnik:1988am,Volkov:1998cc}).
In the case of fractional-spin fields it is also possible to obtain various spherically symmetric systems consisting of gravitating linear
 \cite{Finster:1998ws,Finster:1998ux,Herdeiro:2017fhv}  and nonlinear spin-1/2 fields~\cite{Krechet:2014nda,Adanhounme:2012cm,Bronnikov:2019nqa}.
Nonlinear spinor fields are also employed in studying cylindrically symmetric solutions~\cite{Bronnikov:2004uu}, wormhole solutions~\cite{Bronnikov:2009na},
and various cosmological problems (see Refs.~\cite{Ribas:2010zj,Ribas:2016ulz,Saha:2016cbu} and references inside).
In turn, the inclusion of rotation in the system enables one to get axially symmetric solutions for configurations supported by linear spin-1 and spin-1/2
 fields~\cite{Herdeiro:2019mbz}.

In our previous works, we have considered static gravitating configurations supported by two nonlinear spin-1/2 fields. In doing so,
the cases both of purely Einstein-Dirac systems~\cite{Dzhunushaliev:2018jhj} and of configurations containing also U(1) Maxwell and Proca fields~\cite{Dzhunushaliev:2019kiy},
 or SU(2) Yang-Mills and Proca fields~\cite{Dzhunushaliev:2019uft} have been investigated. The remarkable feature of all such systems is that,
 for some values of the spinor field coupling constant $\lambda$,
one can obtain configurations whose masses are comparable to the Chandrasekhar mass and effective radii are of the order of
kilometers. Such objects may already be regarded as self-gravitating Dirac stars which are prevented from collapse
under their own gravitational fields due to the Heisenberg uncertainty principle.

In the next step, in Ref.~\cite{Dzhunushaliev:2021ztz}, we have used the fact
that  for physically reasonable values of the coupling constant~$\lambda$, it is possible to
replace the nonlinear spinor fields by a hydrodynamic fluid described by an effective equation of state (EOS). This allowed us, within Newtonian gravity, to consider
static compact objects and to study a dynamical process of their creation through the gravitational collapse of such a fluid.

The next natural step in studying such type of configurations is the inclusion of rotation in the system; this will be done in the present paper.
To do this, as in Ref.~\cite{Dzhunushaliev:2021ztz}, we will employ the effective EOS that describes well configurations supported by the nonlinear spinor field in the physically interesting
range of system parameters and whose usage does not require a consideration of  a much more complicated problem
when the nonlinear spinor field is directly involved. We will work within Einstein's general relativity and study rapidly rotating configurations which are in their own strong gravitational field.

The paper is organized as follows.  In Sec.~\ref{spinor_fluid}, we introduce the effective equation of state for a spinor fluid.
In Sec.~\ref{gen_eqs}, within general relativity, we write down the general field equations sourced by the spinor fluid.
In Sec.~\ref{num_sol}, we solve these equations numerically and
obtain spherically and axially symmetric solutions describing compact static and rapidly rotating configurations.
Finally, in Sec.~\ref{conclus}, we summarize the results obtained.

\section{Equation of state for a spinor fluid}
\label{spinor_fluid}

 In Refs.~\cite{Dzhunushaliev:2018jhj,Dzhunushaliev:2019kiy,Dzhunushaliev:2019uft} we have studied gravitating systems with a doublet of a nonlinear spin-1/2 field $\psi$
 described by the Lagrangian
\begin{equation}
\label{Lagran}
	L_{\text{sp}} =	\frac{i \hbar c}{2} \left(
			\bar \psi \gamma^\mu \psi_{; \mu} -
			\bar \psi_{; \mu} \gamma^\mu \psi
		\right) - \mu c^2 \bar \psi \psi + \frac{\lambda}{2} \left(\bar\psi\psi\right)^2,
\end{equation}
where $\lambda$  is the coupling constant and $\mu$ is the mass of the spinor field.

In Ref.~\cite{Dzhunushaliev:2018jhj}, it was demonstrated that the introduction of the dimensionless coupling constant
 $\bar \lambda \sim \left(M_{\text{Pl}}/\mu\right)^2\lambda$, where $M_{\text{Pl}}$ is the Planck mass,
 in the limit of $|\bar \lambda|\gg 1$, permits one to get configurations whose sizes and masses are comparable to those typical
 of neutron stars.
 It was also shown there that such limiting configurations can be described by an effective EOS relating the pressure
$p$  and the energy density $\varepsilon$ as
\begin{equation}
\label{EoS_eff}
	p = \frac{\varepsilon_0}{9} \left(
		1 + 3\frac{\varepsilon}{\varepsilon_0}-\sqrt{1 + 6\frac{\varepsilon}{\varepsilon_0}}
	\right),
\end{equation}
where $\varepsilon_0=\mu c^2/\lambda_c^3\equiv\mu^4 c^5/\hbar^3$ can be regarded as a characteristic
energy density of the system [here $\lambda_c=\hbar/(\mu c)$ is the constant, which
need not be associated with the Compton length since we consider a classical theory].
A configuration described by such an effective EOS can then be regarded as consisting of a fluid, which will be further called {\it a spinor fluid}.

It is worth emphasizing here that in the general case of configurations supported by spinor fields, there are two different effective pressures~--
 the radial, $p_r$, and tangential, $p_t$. Correspondingly, the system is in general anisotropic~\cite{Dzhunushaliev:2018jhj}.
 However, making use of the results of Ref.~\cite{Dzhunushaliev:2018jhj}, it can be shown that in the limit $|\bar \lambda|\gg 1$
 these two pressures become equal, i.e., the system becomes isotropic. For this reason, we consider here only one
 effective pressure $p=p_r=p_t$ appearing in Eq.~\eqref{EoS_eff}.

\section{General equations}
\label{gen_eqs}

Within the framework of Einstein's general relativity,
we consider compact gravitating configurations consisting of a perfect spinor  fluid described by the EOS~\eqref{EoS_eff}.
The corresponding total action for such a system can be represented in the form
[the metric signature is $(+,-,-,-)$]
\begin{equation}
\label{action_gen}
	S_{\text{tot}} = - \frac{c^3}{16\pi G}\int d^4 x
		\sqrt{-g} R +S_{\text{matter}},
\end{equation}
where $G$ is the Newtonian gravitational constant,
$R$ is the scalar curvature, and $S_{\text{matter}}$ denotes the action of the spinor fluid.

 Varying the action \eqref{action_gen} with respect to the metric, we derive the Einstein equations
 \begin{equation}
\label{Ein_gen}
E_{\mu}^\nu	\equiv R_{\mu}^\nu - \frac{1}{2} \delta_{\mu }^\nu R -
	\frac{8\pi G}{c^4} T_{\mu }^\nu=0
\end{equation}
with the energy-momentum tensor
$
T_{\mu }^\nu=\left(\varepsilon+p\right) U_{\mu } U^\nu -\delta_{\mu }^\nu p,
$
where  $U_{\mu }$ represents the four-velocity of the fluid.

In order to obtain rotating configurations, we use the Lewis-Papapetrou line element for a stationary, axially
symmetric spacetime with two Killing vector fields $\xi=\partial_t, \eta=\partial_\varphi$
in a system of spherical coordinates $\{t, r, \theta, \varphi\}$. In these
coordinates the metric is independent of $t$ and $\varphi$, and can
be expressed using isotropic coordinates as~\cite{Kleihaus:2005me}
\begin{equation}
\label{metric}
ds^2=f (d x^0)^2-\frac{l}{f}\left[g\left(dr^2+r^2d\theta^2\right)+r^2\sin^2\theta\left(d\varphi-\frac{\omega}{r}dx^0\right)^2\right],
\end{equation}
where the metric functions $f,l,g$, and $\omega$ depend solely on $r$ and $\theta$, and $x^0=c t$.
The $z$-axis ($\theta=0$) represents the symmetry axis of the system.
Asymptotically (as $r\to \infty$), the functions $f, l, g \to 1$ and $\omega \to 0$; i.e., the spacetime approaches Minkowski spacetime.

We consider here uniform (or rigid) rotation of the spinor fluid; such an assumption is well justified, e.g., for most neutron stars~\cite{HPY}.
In this case the four-velocity is of the form
\begin{equation}
\label{four_vel}
U^\mu=\left(u,0,0,\frac{\Omega}{c}u\right),
\end{equation}
where $\Omega$ is a constant parameter denoting the angular velocity of the system.

For numerical calculations, it is convenient to introduce the dimensionless variables
\begin{equation}
\label{dmls_var}
x=\sqrt{8\pi}\frac{r}{R_L},  \quad \left(\tilde\varepsilon, \tilde p\right)=\frac{\left(\varepsilon, p\right)}{\varepsilon_0}, \quad \tilde \Omega=\frac{R_L}{\sqrt{8\pi}c}\Omega
\quad \text{with}  \quad R_L=\lambda_c\frac{M_{\text{Pl}}}{\mu}.
\end{equation}
The parameter $R_L$ is the characteristic radius obtained by
Landau in considering compact configurations supported by
an ultrarelativistic degenerate Fermi gas and modeled within the framework of Newtonian gravity.

Next, taking into account the condition for the four-velocity $U^\mu U_\mu=1$
and making use of the metric \eqref{metric} and the expression \eqref{four_vel}, the velocity $u$
can be expressed in terms of the metric functions $f, l$, and $\omega$ as follows [written already in terms of the dimensionless variables~\eqref{dmls_var}]:
\begin{equation}
\label{expres_u}
u^2=\frac{1}{f\left[1-x^2 f^{-2} l \sin^2\theta\left(\tilde\Omega-\omega/x\right)^2\right]}.
\end{equation}

In turn, the conservation law $\nabla_{\mu}T^{\mu\nu}=0$ yields the differential equations
$$\frac{\partial_r p}{\varepsilon+p}=\frac{\partial_r u}{u},\quad \frac{\partial_\theta p}{\varepsilon+p}=\frac{\partial_\theta u}{u}.$$
Substituting in these equations the EOS~\eqref{EoS_eff} and integrating them, one can find the expression for the energy density in terms of the four-velocity,
$$\tilde \varepsilon= \frac{1}{32}\left(3 c_0^4u^4-4 c_0^2 u^2-4\right),$$
where $c_0$ is an integration constant. In turn, substituting here the expression~\eqref{expres_u}, one can get
\begin{equation}
\label{expres_eps_2}
\tilde \varepsilon=\frac{1}{24}\left\{
A\frac{f_c f}{f^2-l\sin^2\theta\left(\omega-\tilde\Omega x \right)^2}\left[
A\frac{f_c f}{f^2-l\sin^2\theta\left(\omega-\tilde\Omega x \right)^2}-2
\right]-3
\right\}.
\end{equation}
For convenience, we have introduced here a new constant $A$ defined by the relation $c_0^2=2/3 A f_c$, where $f_c$ is the central value of the metric function $f$.

The field equations coming from the Einstein equations \eqref{Ein_gen} can be represented in the following form [written in terms of the dimensionless variables~\eqref{dmls_var}]:
\begin{align}
&f_{xx}+\frac{l \sin^2\theta}{f}\left(\frac{2}{x}\omega-\omega_x\right)\omega_x+\left(\frac{2}{x}+\frac{l_x}{2 l}\right)f_x-
\frac{f_x^2}{f}+\frac{1}{x^2}\left(
f_{\theta\theta}-\frac{l\sin^2\theta}{ f}\omega_{\theta}^2+\frac{f_\theta l_\theta}{2  l}-
\frac{f_\theta^2}{ f}+\cot\theta f_\theta-\frac{l\sin^2\theta\omega^2}{f}\right)\nonumber\\
&=
\frac{g l}{f}\left\{\left(\tilde\varepsilon+\tilde p\right)\left[f^2+l\sin^2\theta\left(\omega-\tilde\Omega x\right)^2
\right]u^2+2 f \tilde p
\right\},
\label{eq_f}\\
&l_{xx}-\frac{l_x^2}{2 l}+\frac{3 l_x}{x}+\frac{1}{x^2}\left(l_{\theta\theta}-\frac{l_\theta^2}{2 l}+2\cot\theta l_\theta
\right)=4\frac{g l^2}{f}\tilde p,
\label{eq_l}\\
&g_{xx}+\frac{3 g l\sin^2\theta}{f^2}\left(\frac{\omega}{x}-\frac{\omega_x}{2}\right)\omega_x-\frac{g}{l}\left(\frac{2}{x}+\frac{l_x}{2 l}\right)l_x-
\frac{g_x^2}{g}+\frac{g_x}{x}+\frac{g f_x^2}{2 f^2}
\nonumber\\
&+\frac{1}{x^2}\left(g_{\theta\theta}-\frac{3 g l\sin^2\theta\omega_\theta^2}{2 f^2}-
\frac{g l_\theta^2}{2 l^2}-\frac{2\cot\theta g l_\theta}{l}-\frac{g_\theta^2}{g}+\frac{g f_\theta^2}{2 f^2}-\frac{3 g l\sin^2\theta \omega^2}{2 f^2}
\right)
\nonumber\\
&=\frac{2 g^2 l}{f^2}\left[ u^2  l\sin^2\theta \left(\tilde\varepsilon+\tilde p\right)\left(\omega- \tilde\Omega x\right)^2-f\tilde p
\right],
\label{eq_g}\\
&\omega_{xx}+\left(\frac{2}{x}-\frac{2 f_x}{f}+\frac{3 l_x}{2 l}\right)\omega_x-\frac{3\omega l_x}{2 x l}+\frac{2\omega f_x}{x f}+
\frac{1}{x^2}\left[\omega_{\theta\theta}+\left(3\cot\theta-\frac{2 f_\theta}{f}+\frac{3 l_\theta}{2 l}
\right)\omega_\theta-2\omega
\right]=2  u^2 g l \left(\tilde\varepsilon+\tilde p\right)\left(\omega-\tilde\Omega x\right).
\label{eq_omega}
\end{align}
(Here the lower indices denote differentiation with respect to the corresponding coordinate.)
These equations are the following combinations of the components of the Einstein equations~\eqref{Ein_gen}:
$\left(E_t^t-E_r^r-E_\theta^\theta-E_\varphi^\varphi+2\,\omega/r E_\varphi^t\right)=0$,
$\left(E_r^r+E_\theta^\theta\right)=0$, $\left(E_\varphi^\varphi-E_r^r-E_\theta^\theta-\omega/r E_\varphi^t\right)=0$,
and $E_\varphi^t=0$.

Apart from the Einstein equations~\eqref{eq_f}-\eqref{eq_omega}, which are elliptic partial differential equations, there are also two more gravitational equations $E^r_\theta=0$ and
$\left(E^r_r-E^\theta_\theta\right)=0$, whose structure is not of Laplace form and which can be regarded as ``constraints''.
According to arguments proposed in Ref.~\cite{Wiseman:2002zc}, to find a self-consistent solution to
Eqs.~\eqref{eq_f}-\eqref{eq_omega}, it is enough to check that the constraint $E^r_\theta=0$ is satisfied on all boundaries;
this in turn implies that it is satisfied throughout the region of integration.
Then the constraint $\left(E^r_r-E^\theta_\theta\right)=0$ must only be imposed at a single point to ensure that it is fulfilled throughout the region of integration as well.
Notice that, as applied to Dirac stars supported by a linear spinor field, such an approach to obtaining
self-consistent solutions has been used in Ref.~\cite{Herdeiro:2019mbz}. In the present paper, in order to find self-consistent numerical solutions, we will also
check whether these constraints are fulfilled.

\section{Numerical solutions}
\label{num_sol}
In this section we obtain sequences of static and rotating solutions to Eqs.~\eqref{eq_f}-\eqref{eq_omega} for the whole physically relevant range of system parameters.

\subsection{Boundary conditions}

We will seek globally regular, asymptotically flat solutions describing configurations with a finite mass. For such solutions, we impose
appropriate boundary conditions for the metric functions at the origin ($x=0$), at infinity ($x\to \infty$),
on the positive $z$-axis ($\theta=0$), and, making use of the reflection symmetry with respect to $\theta\to \pi-\theta$,
in the equatorial plane ($\theta=\pi/2$). So we require
\begin{align}
\label{BCs}
\begin{split}
&\left. \frac{\partial f}{\partial x}\right|_{x = 0} =
\left. \frac{\partial l}{\partial x}\right|_{x = 0} =  0,  \left. g \right|_{x = 0} = 1, \left. \omega \right|_{x = 0} = 0;\\
&\left. f \right|_{x = \infty} = \left. g \right|_{x = \infty} =\left. l \right|_{x = \infty} =1 ,
	\left. \omega \right|_{x = \infty} = 0  ; \\
&\left. \frac{\partial f}{\partial \theta}\right|_{\theta = 0,\pi} =
\left. \frac{\partial l}{\partial \theta}\right|_{\theta = 0,\pi} =
	\left. \frac{\partial \omega}{\partial \theta}\right|_{\theta = 0,\pi} =  0 ,  \left. g \right|_{\theta = 0,\pi} = 1 ;\\
&\left. \frac{\partial f}{\partial \theta}\right|_{\theta = \pi/2} =\left. \frac{\partial g}{\partial \theta}\right|_{\theta = \pi/2} =
\left. \frac{\partial l}{\partial \theta}\right|_{\theta = \pi/2} =
	\left. \frac{\partial \omega}{\partial \theta}\right|_{\theta = \pi/2}  =  0.
\end{split}
\end{align}
Notice that the elementary flatness condition implies that on the symmetry axis the function $g|_{\theta=0,\pi}=1$ \cite{Kleihaus:2005me};
this ensures the absence of a conical singularity.

\subsection{Asymptotic behavior}

Since we seek asymptotically flat solutions, we require the metric to approach the Minkowski
metric at spatial infinity, i.e.,  asymptotically, $f,g,l\to 1$ and $\omega\to 0$. For our purposes, let us write down expressions describing an asymptotic
behavior of the metric functions
$$
f\approx 1-\frac{2 G M}{c^2 r}+{\cal O}\left(r^{-2}\right),\quad \omega\approx \frac{2G J}{c^3 r^2}+{\cal O}\left(r^{-3}\right).
$$
These expressions contain the total mass of the star $M$ and the angular momentum $J$, which can be represented
 in the form~\cite{Kleihaus:2005me}
\begin{equation}
\label{expres_mass_mom}
M=\frac{c^2}{2G}\lim_{r\to\infty}r^2\partial_r f, \quad J=\frac{c^3}{2G}\lim_{r\to\infty}r^2\omega.
\end{equation}

\subsection{Center and surface of the star}

Let us now write out expressions for central values of the energy density and pressure of the star. Using Eqs.~\eqref{EoS_eff} and
\eqref{expres_eps_2} and taking into account that at the centre $\left. \omega \right|_{x = 0} = 0$ [see the boundary conditions~\eqref{BCs}],
one can obtain
$$
\tilde \varepsilon_c=\frac{1}{24}\left[A^2 -2 A-3\right], \quad \tilde p_c=\frac{1}{9}\left(1+3\tilde \varepsilon_c-\sqrt{1+6\tilde \varepsilon_c}\right).
$$
By choosing the value of the arbitrary constant $A=1+2\sqrt{7}$, we have $\tilde \varepsilon_c=1$. In turn, given different values of $A$,
one can construct families of solutions describing configurations with different physical characteristics (see below).

The boundary of the star is defined as the surface with radius $r_b(\theta)$ where $\varepsilon\to 0$ (correspondingly, $p \to 0$ as well).
To construct an embedding diagram that depicts the intrinsic geometry of the stellar surface [i.e., the slice of spacetime with $t=\text{const.}$ and $r=r_b(\theta)$],
one can embed the stellar surface in a flat three-dimensional space~\cite{Friedman:1986tx}.
The metric of the stellar surface induced by the spacetime metric \eqref{metric} is
$$
ds_b^2=\frac{l_b}{f_b}\left\{g_b\left[\left(\partial_\theta r_b\right)^2+r_b^2
\right]d\theta^2+r_b^2\sin^2\theta d\varphi^2
\right\},
$$
where the index $b$ denotes the boundary of the star, on which the metric functions are given [for example, $f_b\equiv f\left(r_b(\theta),\theta\right)$].
Then, changing to cylindrical coordinates $\{\rho, z, \varphi\}$ for the flat space, one can obtain the following expressions~\cite{Friedman:1986tx}:
$$\rho(\theta)=\sqrt{\frac{l_b}{f_b}} r_b \sin\theta, \quad z(\theta)=\int_{\theta}^{\pi/2}d\theta^\prime
\left\{\frac{g_b l_b}{f_b}\left[\left(\frac{d r_b}{d\theta^\prime}\right)^2+r_b^2\right]-\left(\frac{d\rho}{d\theta^\prime}\right)^2
\right\}^{1/2}.$$

Making use of them, the equatorial and polar radii of the embedded surface are then given by
\begin{align}
&R_e=\rho\left(\theta=\frac{\pi}{2}\right)=\left.\sqrt{\frac{l_b}{f_b}} r_b\right|_{\theta = \pi/2} ,
\label{rad_eq}
\\
&R_p=z(\theta=0)=\int_{0}^{\pi/2}d\theta^\prime
\left\{\frac{g_b l_b}{f_b}\left[\left(\frac{d r_b}{d\theta^\prime}\right)^2+r_b^2\right]-\left(\frac{d\rho}{d\theta^\prime}\right)^2
\right\}^{1/2}.
\label{rad_pol}
\end{align}

Consider now the question of a limiting angular velocity of rotation (or the Keplerian limit) of the star under consideration.
This limit is reached when the star rotates so fast that the angular velocity $\Omega$ of the fluid approaches the
angular velocity $\Omega_p$ of a free particle located at the equator of the stellar surface.
As a result,  a fluid element cannot already be confined on the surface of the star and this ultimately leads to losing mass.

The corresponding expression for $\Omega_p$ can be found from the geodesic equation for the angular
velocity of the particle~\cite{Kleihaus:2016dui}
\begin{equation}
\label{expres_Op}
\left(\frac{\Omega_p}{c}-\frac{\omega_b}{r}\right)^2-2 a_p\left(\frac{\Omega_p}{c}-\frac{\omega_b}{r}\right)+b_p=0,
\end{equation}
where $\omega_b\equiv \omega(r_b,\pi/2)$ and
$$
a_p= \left. \frac{f l\left(r\partial_r\omega-\omega\right)}{r\left[f\left(2 l+r \partial_r l\right)-r l \partial_r f\right]}\right|_{r_b,\pi/2} ,\quad
b_p= \left. -\frac{f^2 \partial_r f}{r\left[f\left(2 l+r \partial_r l\right)-r l \partial_r f\right]}\right|_{r_b,\pi/2} .
$$
Solving Eq.~\eqref{expres_Op} for $\Omega_p$, one can obtain
$$\frac{\Omega_p}{c}=\frac{\omega_b}{r}+a_p+\sqrt{a_p^2-b_p}.$$
The Keplerian angular velocity $\Omega_K$ is defined as $\Omega_K=\Omega_p$. A necessary condition for the existence
of a stationary rotating configuration implies that the equatorial velocity of stellar matter is smaller than the Keplerian velocity
of a particle moving along the $\varphi$-direction in the equatorial plane. The Keplerian velocity corresponds to the Keplerian angular velocity $\Omega_K$,
also called the mass-shedding angular velocity.

\subsection{Numerical approach}

The set of four coupled nonlinear elliptic partial differential equations~\eqref{eq_f}-\eqref{eq_omega} for the unknown functions $f, g, l$, and $\omega$
will be solved numerically subject to the  boundary conditions~\eqref{BCs}.
Keeping in mind that we will seek even parity solutions which are symmetrical about the equatorial plane $\theta=\pi/2$, numerical computations will be carried out only in the region $0\leq \theta \leq \pi/2$.
In doing so,  for numerical calculations, it is convenient to introduce new compactified coordinate
\begin{equation}
	\bar x=\frac{x}{1+x},
	\label{comp_coord}
\end{equation}
the use of which permits one to map the infinite region $[0,\infty)$ to the finite interval $[0,1]$.

All results of numerical calculations for axially symmetric systems exhibited below have been obtained using the package FIDISOL~\cite{fidisol}
where the numerical procedure based on the Newton-Raphson method is employed.  This method provides  an iterative procedure for getting
an exact solution starting from an approximate solution (an initial guess). As such initial guess, we use the solutions for nonrotating (i.e., static) spherically symmetric configurations
which can be obtained using Eqs.~\eqref{eq_f}-\eqref{eq_omega} with $g=1$  and $\Omega, \omega=0$. In this case these equations
reduce to the set of two ordinary differential equations for the metric functions $f$ and $l$ which have been solved using the NDSolve routine from {\it Mathematica}.

In turn, solutions to the partial differential equations~\eqref{eq_f}-\eqref{eq_omega} are sought on a grid of $401\times 101$ points,
covering  the region of integration $0\leq \bar x \leq 1$ [given by the compactified coordinate from Eq.~\eqref{comp_coord}] and $0\leq \theta \leq \pi/2$.
In performing calculations, we have kept track of the behavior of the energy density at every point in space $(r, \theta)$
in order to determine the position of the boundary of the star $r_b(\theta)$ where $ \varepsilon \to 0$. The case of
$ \varepsilon < 0$ corresponds to the fact that the meshpoint is outside the star; in this case the right-hand sides of Eqs.~\eqref{eq_f}-\eqref{eq_omega}
are taken to be zero (i.e., the vacuum Einstein equations are solved).

\subsection{Static and rotating solutions}

Using the approach described in the previous subsection, let us obtain sequences of solutions describing Dirac stars.
The strategy of their obtaining is as follows. First, as seed configurations, we obtain spherically symmetric, nonrotating configurations,
starting the solution in the neighborhood of the center where the metric functions behave as
$$
f\approx f_c+\frac{1}{2}f_2 x^2+ \ldots, \quad l\approx l_c+\frac{1}{2}l_2 x^2+ \ldots
$$
In doing so, to ensure the asymptotic flatness of the spacetime, it is necessary to take strictly defined central values $f_c$ and $l_c$
(eigenvalue problem) which ensure $f, l \to 1$ at infinity. Then, by varying the integration constant $A$ appearing in the expression for the energy density
\eqref{expres_eps_2}, one can find the family of static configurations with $\omega=\Omega=0$ possessing different masses and radii of the spinor fluid, which are
parameterized by $A$.

\begin{figure}[t]
\includegraphics[width=1\linewidth]{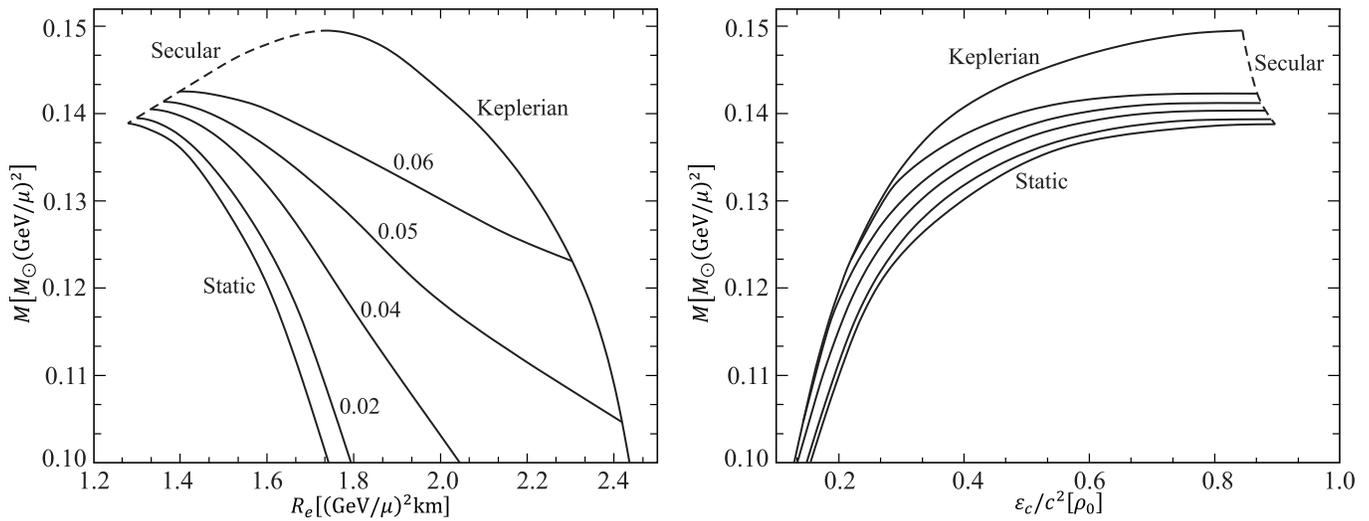}
\vspace{-0.5cm}
\caption{
Left panel: the mass-radius relation for Dirac stars in the physically relevant domain for different values of the dimensionless
angular velocity $\tilde \Omega$ (designated by the numbers near the curves). For $\mu= 0.3~\text{GeV}$, the value  $\tilde \Omega=0.01$
corresponds to a frequency of $f\approx 90\, \text{Hz}$.
Right panel: the mass-central energy density relation for the systems of the left panel [the energy density is expressed in units of $\rho_0$ from Eq.~\eqref{rho0}].
}
\label{fig_Mass_rad_dens}
\end{figure}

Next, using the static systems as an initial guess, we gradually increase the velocity of rotation ($\Omega\neq 0$) and keep track of changes of characteristics
of the stars under consideration. The results of calculations are given in Fig.~\ref{fig_Mass_rad_dens}, where the left panel shows the mass-radius relation for the systems
with different velocities of rotation. To plot these dependences, we have used the following  dimensional quantities
 [obtained using Eqs.~\eqref{dmls_var} and \eqref{expres_mass_mom}]:
\begin{equation}
\label{mass_dim}
r_b(\theta)\approx 0.48 \frac{\bar x_b}{1-\bar x_b}\left(\frac{\text{GeV}}{\mu}\right)^2\,\text{km}, \quad
M\approx 0.1623\lim_{\bar x\to 1}\bar x^2\frac{\partial f}{\partial \bar x} M_{\odot}\left(\frac{\text{GeV}}{\mu}\right)^2  .
\end{equation}
To find the equatorial radius $R_e$, this $r_b(\theta)$ is substituted in Eq.~\eqref{rad_eq}.
In turn, the spin frequency of the configurations under consideration is
$$
f\equiv\frac{\Omega}{2\pi}\approx10^5\tilde \Omega\left(\frac{\mu}{\text{GeV}}\right)^2\,\text{Hz}.
$$

The domain of Dirac stars shown in Fig.~\ref{fig_Mass_rad_dens} is bounded according to the following physical reasons:
\begin{itemize}
\item[(i)] The left boundary of the domain corresponds to the set of static stars whose mass increases monotonically with decreasing the radius up to its maximum.
This maximum corresponds to a point that separates stable configurations (located to the right of the maximum) and unstable ones (located to the left of the maximum,
not shown here).
\item[(ii)] The right boundary of the domain corresponds to the set of configurations rotating at the Keplerian limit, $\Omega=\Omega_K$.
On this boundary the mass also increases monotonically with decreasing the radius up to its maximum that separates stable and unstable systems.
\item[(iii)] The two aforementioned boundaries
are connected at the extreme points by the so-called secular instability line (shown by the dashed line) which is the locus of
the points corresponding to maxima of masses of the configurations rotating with different fixed values of the angular velocity $\tilde \Omega$.
Configurations located to the left of this boundary are unstable (see below in Sec.~\ref{stability}).
\end{itemize}

Technically, the process of obtaining the curve corresponding to the configurations rotating at the Keplerian limit consists in successive increasing the value of
$\Omega$  up to the point where $\Omega\approx\Omega_p$. Since the numerical calculations do not permit us to continue the solution
to the point $\Omega=\Omega_p$, the physical parameters of the systems rotating at the Keplerian limit
have been obtained by their extrapolation
as functions of $\Omega$ in the limit $\Omega\to\Omega_p$~\cite{Kleihaus:2016dui}.

Notice that the above behavior of the mass-radius curves of the Dirac stars under consideration is similar to the behavior of analogous curves typical of neutron stars
(see, e.g., Refs.~\cite{Kleihaus:2016dui,Cipolletta:2015nga}). In particular, by choosing the mass of the spinor field  $\mu= 0.3~\text{GeV}$,
total masses and sizes of the Dirac stars are comparable to characteristics typical of neutron stars.

\begin{figure}[t]
\includegraphics[width=1\linewidth]{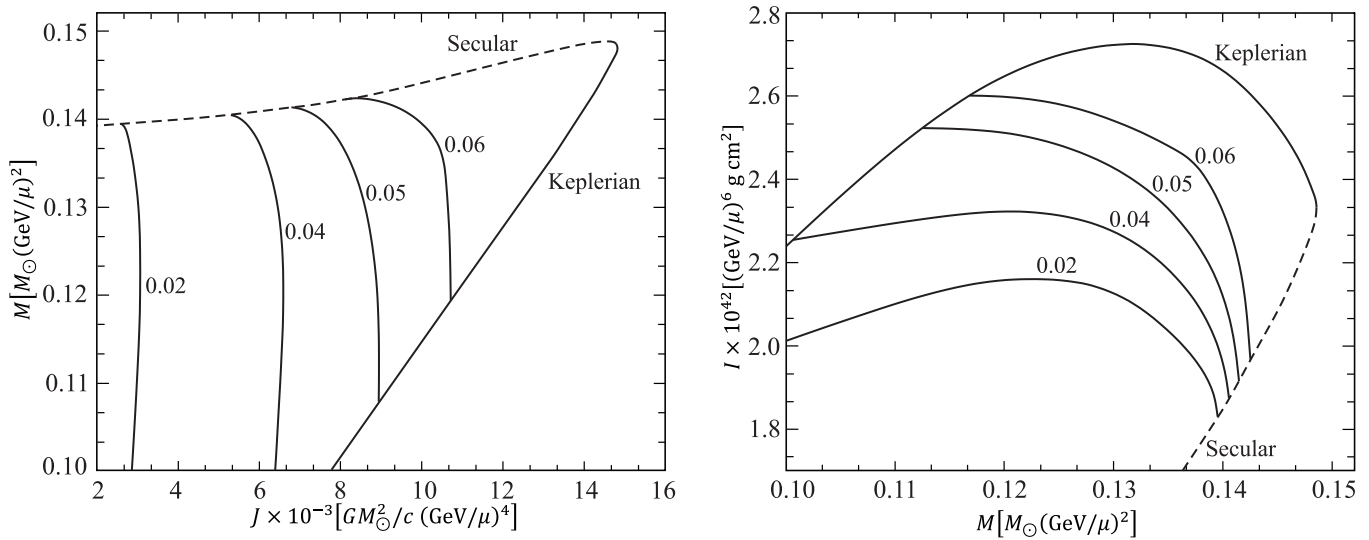}
\vspace{-0.5cm}
\caption{The mass $M$ vs. the angular momentum $J$ (left panel) and the moment of inertia $I$ vs. the mass $M$ (right panel)
are shown for Dirac stars in the physically relevant domain for different values of the dimensionless angular velocity $\tilde \Omega$ (designated by the numbers near the curves).
}
\label{fig_moment}
\end{figure}

\subsection{Stability}
\label{stability}

Let us now address the question of stability of the systems under consideration. In doing so,
we will use the turning-point method of Ref.~\cite{Friedman:1988er} which provides a simple criterion to locate
points of secular instability along a sequence of equilibrium configurations. This method does not require the use of perturbation analysis
but instead deals with a consideration of one- or many-parameter families of solutions.

Namely, in the case of static systems one can estimate stability by considering
the behavior of a mass-central density curve. The maximum mass of a stable configuration coincides with the first maximum of a sequence of
configurations with increasing central density; this takes place at the first point where $\partial M/\partial\varepsilon_c=0$ (the turning point).
As the central density increases further (after it has passed the turning point), there occurs instability that evolves on a secular timescale, i.e., it is not dynamical~\cite{Friedman:1988er}.

When the rotation is turned on, apart from the central density, one more extra parameter is involved; this can be the ratio of the coordinate radii $r_p/r_e$,
the angular velocity $\Omega$, the angular momentum $J$, the gravitational mass $M$. In this case it is always possible to construct a set of rotating configurations
by fixing a value of the second parameter and letting the central energy density vary in a range which is
constrained by stability limits~\cite{Cipolletta:2015nga}. As shown in Ref.~\cite{Friedman:1988er}, the turning-point
method can be applied to uniformly rotating stars. For a sequence of configurations with a constant angular velocity $\Omega$,
the turning point of the sequence with increasing central density serves as a boundary between secularly stable and unstable configurations;
that is, a secular axisymmetric instability occurs at the point where $\left.\partial M(\varepsilon_c, \Omega)/\partial \varepsilon_c\right|_{\Omega=\text{const.}}=0$.
Then a curve connecting all the maxima (i.e., the turning points obtained for different fixed values of $\Omega$)
bounds the stability domain
(cf. the secular instability line shown by the dashed curve in the left panel of Fig.~\ref{fig_Mass_rad_dens}).
The point of intersection of such a limiting curve with the Keplerian sequence corresponds to a configuration rotating with the largest possible angular velocity.

 It is worth mentioning here that, as was pointed out in Ref.~\cite{Takami:2011zc}, the turning point
criterion is only a {\it sufficient} condition for secular instability of rotating stars, and
an analysis of the dynamical stability reveals that dynamical instability occurs for smaller central densities.
Nevertheless,
the turning point method employed here, while being technically more simple, ensures sufficient accuracy
in determining a stability domain.

To illustrate the above for the systems under investigation, consider
the mass-central energy density relation. For greater clarity, instead of the energy density $\varepsilon$, we will use
the mass density $\rho$ and introduce the characteristic mass density
\begin{equation}
\label{rho0}
\rho_0\equiv \frac{\varepsilon_0}{c^2}\approx 2.33\times 10^{17}\left(\frac{\mu}{\text{GeV}}\right)^4 \,\text{g cm}^{-3}
\end{equation}
[the expression for $\varepsilon_0$ can be found after Eq.~\eqref{EoS_eff}]. The corresponding dependences of the mass of the systems under consideration on
the central density are shown in the right panel of Fig.~\ref{fig_Mass_rad_dens}.
The behavior of these dependences is qualitatively similar to the behavior
of the relations obtained for neutron stars: the central density is
maximum for static Dirac stars in the stability limit and decreases with increasing angular velocity while moving along the secular instability line (cf. Refs.~\cite{Kleihaus:2016dui,Cipolletta:2015nga}).
The maximally rotating stable configurations are located on the Keplerian (or mass-shedding) sequence; here the gravitational force equals
the centrifugal force at the star's equator, and a faster rotation would inevitably lead to the expulsion of mass from the star.

To conclude this subsection, notice that if as an example one takes $\mu= 0.3~\text{GeV}$, the characteristic central density of the systems under consideration near the maximum of the mass
is of the order of  $10^{15}\,\text{g cm}^{-3}$;  as in the case of masses and sizes (see the end of the previous subsection), this is the value that is also typical of neutron stars.

\subsection{Angular momentum and moment of inertia}

We turn now to the angular momentum $J$ of rotating Dirac stars. In the left panel of Fig.~\ref{fig_moment}, we exhibit  the dependences of the total mass
 $M$ on the angular momentum for different velocities of rotation. To plot these graphs, we have used the expression for the mass from Eq.~\eqref{mass_dim}
 and the dimensional quantity [obtained using Eqs.~\eqref{dmls_var} and \eqref{expres_mass_mom}]
$$
J\approx 5.27\times 10^{-2}\lim_{\bar x\to 1}\left(\frac{\bar x}{1-\bar x}\right)^2 \omega \,\frac{G M_{\odot}^2}{c}\left(\frac{\text{GeV}}{\mu}\right)^4  .
$$
It is seen from the figure that (i) the secular instability line forms the upper limit for the mass, while the lower limit corresponds to the Keplerian
sequence; and (ii) the relationship between the star's total mass and the angular momentum for the Keplerian sequence is practically linear.

For the sake of comparison, consider the angular momentum of Kerr black holes. For them, the reduced dimensionless
angular momentum
$$
\frac{a}{M}=\frac{c J}{G M^2}
$$
is limited by the maximum value attained by the extremal black holes, $\left(a/M\right)_{\text{max}}=1$.
For neutron stars supported by different EOSs $\left(a/M\right)_{\text{max}}\approx 0.7$~\cite{Cipolletta:2015nga}, i.e., it is considerably lower than the Kerr value.
Also, the quantity $a/M$ varies only slightly with the mass. In turn, in the case of Dirac stars supported by the EOS~\eqref{EoS_eff},
we observe a similar behavior, with the maximum value $\left(a/M\right)_{\text{max}}\approx 0.78$.

Compute now the moment of inertia of the star, which is one of the most relevant properties in analysis of neutron stars (pulsars). The
moment of inertia can be evaluated as
$$
I=\frac{J}{\Omega}.
$$
The right panel of Fig.~\ref{fig_moment} shows $I$ as a function of
the star's mass for several fixed values of the angular velocity $\tilde \Omega$, as well as for the Keplerian sequence.
On comparing these dependences with the results obtained for neutron stars modeled by different EOSs~\cite{Cipolletta:2015nga},
one can observe that there are some qualitative differences in the behavior of the curves, related obviously to the use of another EOS in the form~\eqref{EoS_eff}.
In turn, by choosing as an example  $\mu= 0.3~\text{GeV}$, one can see that $I\sim 10^{45} \text{g cm}^2$;
this is a typical value for neutron stars.

\begin{figure}[t]
\includegraphics[width=1\linewidth]{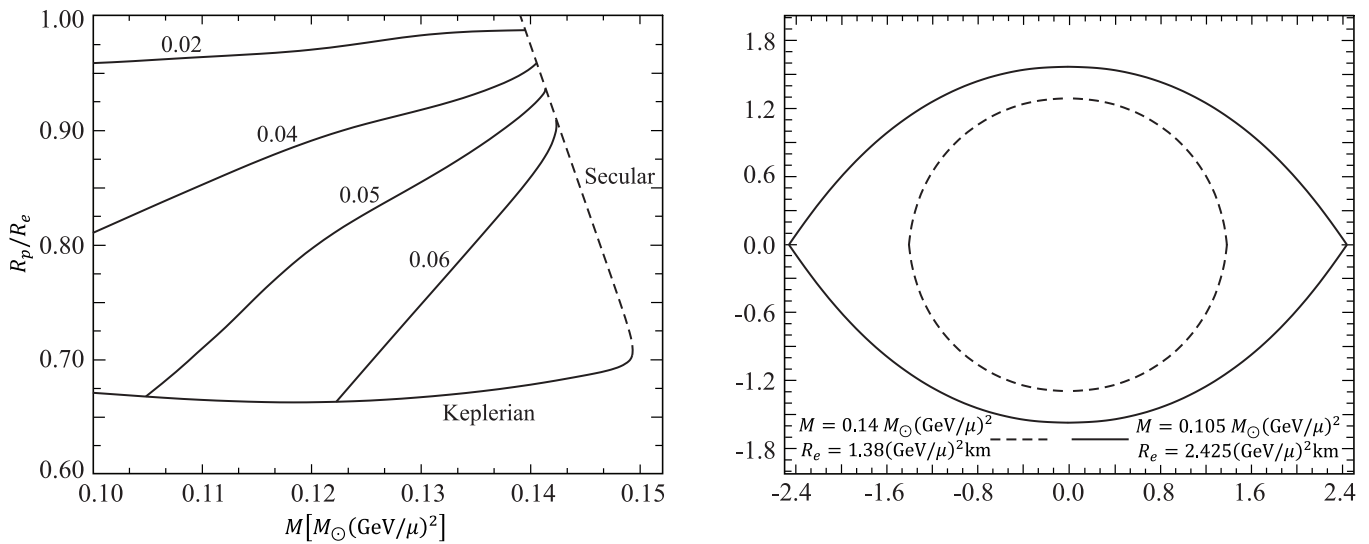}
\vspace{-0.5cm}
\caption{
Left panel: the ratio of the polar to the equatorial radius, $R_p/R_e$, as a function of the total mass $M$ of Dirac stars in the physically
relevant domain for different values of the dimensionless angular velocity $\tilde \Omega$ (designated by the numbers near the curves).
Right panel: the isometric embedding of the surface of a star close to the secular instability line (shown by the dashed curve)
 and of a star close to the Keplerian limit (shown by the solid curve); for both surfaces $\tilde \Omega=0.05$.
}
\label{fig_deform}
\end{figure}

\begin{figure}[t]
\includegraphics[width=1\linewidth]{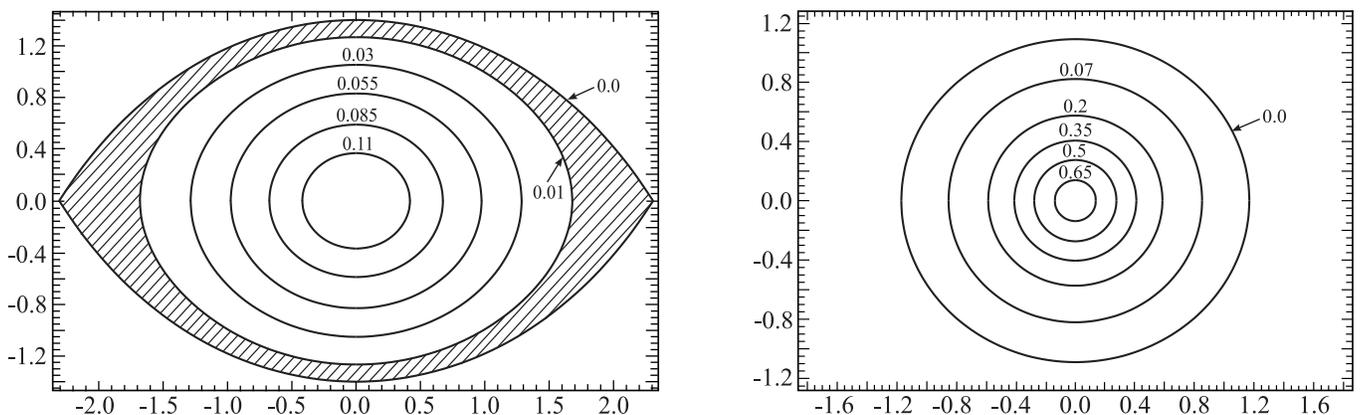}
\vspace{-0.5cm}
\caption{
Isosurfaces of constant energy density $\varepsilon/c^2$ [in units of $\rho_0$ from Eq.~\eqref{rho0}]
for the configurations shown in the right panel of Fig.~\ref{fig_deform}.
The plots are made in a meridional plane $\varphi=\text{const.}$  spanned by the
coordinates $X=r\sin\theta$ and $Z=r\cos\theta$.
The corresponding values of the energy density are designated by the numbers near the curves.
}
\label{fig_dens_distr}
\end{figure}

\subsection{Deformation of the stars}

The presence of rotation results in a deviation of the shape of configurations from spherical symmetry
due to the centrifugal forces which deform the stars.
The flattening of a spinning star can be described by the ratio of the polar radius, $R_p$ to the equatorial radius, $R_e$, defined by
Eqs.~\eqref{rad_pol} and \eqref{rad_eq}, respectively. The left panel of Fig.~\ref{fig_deform}  depicts this
ratio. One can see a rapid increase of the flattening as the angular velocity approaches the Keplerian limit $\Omega_K$.
Another effect of the centrifugal forces is that, at a given $M$, the central density decreases with increasing $\Omega$ (see the right panel of Fig.~\ref{fig_Mass_rad_dens}).

To visualize the above effects, in the right panel of Fig.~\ref{fig_deform}, we show the geometry of the surface of the Dirac stars using the isometric embeddings of the surface for
two configurations: one for the star that is close to the secular instability line and one for the system close to the Keplerian limit.
It is seen that for the stars rotating close to the Keplerian limit, there occurs a cusp at the star's surface  in the equatorial plane.

One more physical quantity of interest is the distribution of the energy density $\varepsilon$. In Fig.~\ref{fig_dens_distr}, we depict contours
of constant energy density.
For illustrative purposes, 
we employ the same Dirac star solutions as those used for constructing the isometric embeddings given in the right panel of Fig.~\ref{fig_deform}.
It is clearly seen from the left panel of Fig.~\ref{fig_dens_distr} that
the centrifugal forces produce the strongest deformation in the equatorial layer of the outer part of the configuration
rotating with the angular velocity close to the Keplerian limit (shown by the shaded region).

 \section{Conclusion}
\label{conclus}

In the present paper, we have been continuing our investigations of the Dirac stars began in Refs.~\cite{Dzhunushaliev:2018jhj,Dzhunushaliev:2019kiy,Dzhunushaliev:2019uft,Dzhunushaliev:2021ztz}.
Our main purpose here is to study the influence of rapid rotation on the characteristics of the Dirac stars.
In order to describe the matter of such stars, we have taken advantage of the fact that, for some values of the spinor field coupling constant $\lambda$
appearing in the Lagrangian~\eqref{Lagran}, it is possible to introduce an effective equation of state in the form~\eqref{EoS_eff}
that adequately describes the system under investigation in the physically interesting range of system parameters.
For the case of uniform rotation considered here, the EOS~\eqref{EoS_eff} permits to integrate analytically  the
differential equations coming from the conservation law and to derive expressions for the energy density and pressure in terms of metric functions.
These expressions have been used as a source of a gravitational field modeled within Einstein's general relativity.
We have started out with a consideration of  static equilibrium solutions describing spherically symmetric Dirac stars.
The inclusion of rotation in such systems results in the fact that the configurations become axisymmetric and, as the velocity of rotation approaches the Keplerian limit,
considerably flattened.

Summarizing the results obtained,
\begin{itemize}
\item[(i)] We have found regular solutions to the Einstein equations with a source in the form of a perfect spinor fluid describing sequences of static and uniformly rotating
Dirac stars.
\item[(ii)] For such configurations, the gravitational mass, equatorial and polar radii, angular momentum, and moment of inertia have been calculated.
\item[(iii)] The mass-radius relations have been constructed, using which we have established the domain of stability which is bounded by the Keplerian limit and secular instability line.
\item[(iv)] The typical distributions of the energy density inside the star and shapes of the objects obtained are exhibited.
\item[(v)] It is demonstrated that for the choice of the spinor field mass $\mu\sim 0.3~\text{GeV}$ the physical characteristics of the Dirac stars
(total masses, radii, densities, velocities of rotation, angular momentum, and moment of inertia) are close to those that are typical of rotating neutron stars.
\end{itemize}

\section*{Acknowledgments}
We would like to thank L.~Rezzolla for discussions and useful suggestions.
This research has been funded by the Science Committee of the Ministry of Education and Science of the Republic of Kazakhstan
(Grant No.~BR10965191 ``Complex research in nuclear and radiation physics, high-energy physics and cosmology for development of the competitive technologies'').
We are also grateful to the Research Group Linkage Programme of the Alexander von Humboldt Foundation for the support of this research.

\end{document}